# Does Adoption of Zero Tillage Reduce Crop Residue Burning? Evidence from Satellite Remote Sensing and Household Survey Data in India

Dominik Naeher*     Virginia Ziulu⁺

Previous research indicates that zero tillage technology offers a profitable alternative to crop residue burning, with significant potential to reduce agricultural emissions and contribute to improvements in air quality and public health. Yet, empirical evidence on the link between zero tillage adoption and residue burning remains scarce, adding to the difficulties policy makers face in this context. This study addresses this gap by integrating high-resolution satellite imagery with household survey data from India to examine the empirical relationship between zero tillage and residue burning. We compare different methods for constructing burn indicators from remote-sensing data and assess their predictive power against survey-based measures. Our findings reveal a robust negative association between zero tillage and crop residue burning, with reductions in the incidence of burning of 50% or more across both survey data and satellite-derived indicators. By providing insights into optimal geospatial data integration methods, our study also makes a methodological contribution that can inform future research and support evidence-based policy interventions for more sustainable agricultural practices.

* Corresponding author. University of Göttingen, Waldweg 26, 37073 Göttingen, Germany (e-mail: dnaeher@uni-goettingen.de).
⁺ World Bank Group, Washington, DC, United States (e-mail: vziulu@worldbank.org).

*Declarations of interest*: The authors declare no competing interests. The views expressed in this paper are solely those of the authors and do not necessarily reflect the views of the World Bank Group.
*Funding*: None.

I. **Introduction**

Crop residue burning is an agricultural practice where farmers intentionally set fire to the straw stubble that remains after harvesting grains like rice and wheat. This practice continues to be widely used across many developing countries, contributing significantly to excessive air pollution, child mortality, and other health burdens (Liu et al. 2020, Lan et al. 2022, Dipoppa & Gulzar 2024).

Zero tillage (or no-till) farming is an agricultural technique for growing crops with minimal disturbance of the soil, including by abstaining from plowing the fields and using seed drill machines to seed wheat directly into unplowed fields (Baeumer & Bakermans 1974, Erenstein & Laxmi 2008).[1] An increasing body of evidence suggests that adoption of zero tillage approaches can be a profitable alternative to crop residue burning, with large potential to reduce both costs and emissions from agricultural land preparation, increase long-term farming productivity, and contribute to improvements in air quality and human health outcomes in affected regions (Schillinger et al. 2010, Mangalassery et al. 2015, Shyamsundar et al. 2019, Keil et al. 2021, Deshpande et al. 2024). Yet, evidence on the empirical relationship between use of zero tillage and residue burning remains scarce, adding to the difficulties policy makers face in assessing the impacts of interventions aimed at reducing crop residue burning (Lan et al. 2022, Dipoppa & Gulzar 2024).

This study addresses this gap by investigating the empirical relationship between zero tillage adoption and crop residue burning in India using a unique dataset constructed by combining high-resolution, publicly available satellite imagery data on crop residue burning with rich information on households' agricultural practices and plot-level variables from recently collected household survey data.

Our contribution to the literature is twofold. First, we compare different methods for constructing burn indicators from geospatial (satellite imagery) data, both among each other and with farmer-reported information on crop residue burning from survey data, generating insights into which methodology for constructing burn indicators from remote-sensing data performs best in matching the survey data.[2] In particular, we investigate whether the availability of plot coordinates provides additional value over household (or village center) coordinates in this context, and which parameter and threshold choices in the construction of remote-sensing burn indicators yield the highest predictive power of outcomes observed in the survey data.

Second, we use the different data sources and constructed burn indicators to investigate the empirical determinants of crop residue burning in a sample of 1,143 households in Northwest India, with a focus on estimating the empirical relationship between the use of zero tillage and instances of crop residue burning.

Our results indicate a robust, negative and statistically significant association between zero tillage and crop residue burning, making zero tillage a strong predictor of stubble burning observed either from farmers' self-reported practices or from remote-sensing data. When focusing on a survey-based measure of residue burning, our estimates indicate that plots that are cultivated with zero tillage are 3.9 percentage points less likely to be subject to crop residue burning. Given a baseline share of 5.5% of plots with residue burning in our study sample, this amounts to a reduction in the likelihood of burning greater than two thirds (70.9%). When using a remote-sensing burn indicator, we find that zero tillage is associated with a reduction in the extent of stubble burning of one half (50.0%).

---

[1] The prevailing zero-tillage technology in our study region in India uses a tractor-drawn seed drill with narrow slits for placing seed and fertilizers into the unplowed soil (Erenstein & Laxmi 2008). Besides helping to keep carbon stored in the soil, zero tillage can also limit soil erosion, reduce water run-off, and allow valuable nutrients and moisture to remain in the soil (Joshi et al. 2007, Schillinger et al. 2010).

[2] This process is sometimes referred to as "on-ground" verification, albeit in our case it is subject to potential misreporting of residue burning in the survey data.

Our study also contributes to the broader literature that seeks to combine geospatial and non-geospatial data sources, a task that continues to pose practical and methodological challenges despite being well recognized in the field. As geospatial data—particularly from high-resolution remote sensing—becomes increasingly available, researchers are often required to integrate it with traditional data sources such as government statistics or household surveys. While this integration is not novel per se, what remains challenging is the operationalization of such integration when variables from non-geospatial data are only observed at discrete point locations (e.g., geo-coded households), while geospatial variables are defined over continuous surfaces or grid cells. Our analysis illustrates one way to address this common challenge, highlighting the implications of spatial aggregation choices on inference.

For instance, consider the common scenario in development and agricultural economics in which one wants to estimate the empirical relationship between (i) agricultural practices observed at the household (or plot) level from survey data, and (ii) geospatial outcomes captured via remote sensing (e.g., satellite imagery). Which grid cells (or parts thereof) should be mapped to each household observation? Is it necessary to have geo-coordinates of the agricultural plots or do household (or village center) coordinates suffice for obtaining good approximations? How should the values for the respective grid cells be aggregated, and how sensitive are the results to these choices?

Our study addresses these questions in the context of crop residue burning in India. This setting is particularly suited for this purpose as the use of remote sensing-derived indicators for quantifying agricultural burns is a well-established practice that we can build on in our investigation (McCarty et al., 2012; Mohammad et al., 2023; Walker et al., 2022; Roy et al., 2019; Deshpande et al., 2022; Walker, 2024; Chandel et al., 2022; Suresh Babu et al., 2018). Our analysis demonstrates that incorporating plot-level coordinates significantly improves the accuracy of remote-sensing burn indicators. In our study, we identify and validate thresholding ratios for spectral indices that proved effective in our setting, offering practical guidance that may be useful in similar contexts. While these thresholds are not universally generalizable, they can serve as a starting point for researchers working in comparable environments, potentially reducing the burden of parameter tuning.

The rest of the paper is organized as follows. Section II lays out the empirical strategy to measure crops residue burning and examine its empirical link to zero tillage adoption. Section III describes the variables and data sources. Section IV documents the construction of remote-sensing burn indicators. Section V presents the results. Section VI concludes.

**II.    Empirical strategy**

*A.   Measuring crop residue burning*

Our analysis considers two distinct approaches to measuring crop residue burning, one based on survey data and one based on remote sensing (satellite imagery) data. An advantage of the survey-based approach is the level of observation, with information on crop residue burning (as well as on zero tillage and other agricultural variables) often available for individual plots. This is an important feature, as estimating the relationship between zero tillage and residue burning will optimally exploit variation at the plot level, as opposed to variation at aggregated levels such as villages.

In contrast, remote-sensing data on residue burning is typically not observed at the plot level, but for larger grid cells which comprise multiple plots (or parts thereof). Even if there was remote-sensing data on residue burning available of sufficient precision for individual plots (which, in principle, might be constructed using visual inspection of high-resolution satellite images to identify plots), mapping this information to data on zero tillage use for each plot would require either (1) remote-sensing data on zero tillage at the plot level, or (2) survey data on both zero tillage use at the plot level and the exact geo-coordinates of each surveyed plot. Regarding (1), the few existing attempts to quantifying the extent of zero tillage in smallholder systems using remote-sensing data have been limited to measuring zero tillage at regional scales, with remote-sensing measures of zero tillage at the plot level not (yet) available (Zhou et al. 2021, Deshpande et al. 2024). A major obstacle to approach (2) is that plot coordinates are typically unavailable in publicly available agricultural household surveys, either because their collection would be too time-consuming and expensive, or because plot geo-coordinates that were collected need to be manipulated to ensure respondents' anonymity, which makes it impossible to identify individual plots and merge them with geocoded variables at the plot level from other datasets.[3]

Thus, unless very-high resolution imagery is used, estimating the relationship between zero tillage and residue burning using remote-sensing data on residue burning is currently only feasible at aggregated levels such as villages, offering less precision (and less statistical power) than plot-level regressions. Moreover, the spatial resolution of publicly available remote sensing indicators is typically not sufficient to capture small fires, and limited temporal resolution (exacerbated by cloud obstructions) may miss to capture short-lived fires. This is less of an issue in survey data, where farmers can be asked whether they performed any burning (no matter how short-lived) during a given period. More broadly, it is important to recognize that remotely sensed data provide powerful observational capabilities but are not devoid of measurement error. Common sources include spatial misalignment, cloud-related gaps, confusion between visually similar surface types, atmospheric interference, and sensor malfunctions (Walker 2024, Lunetta et al. 1991, Povey et al. 2015). These issues go beyond technicalities—they could directly impact the validity of statistical inference.

On the other hand, it should be noted that, unlike remote-sensing, survey data on farmers' self-reported agricultural practices may suffer from misreporting; for example due to imperfect recall, social desirability bias, or data entry errors. This issue is of particular concern for variables with strong social norms or legally binding constraints; thus including the practice of stubble burning, which is both illegal and socially sensitive in India.[4] It is therefore *a priori* not clear which of the two approaches - survey vs. remote-sensing - yields more accurate measures of crop residue burning.

In our empirical investigation, we will focus on comparing the performances of different remote-sensing burn indicators in predicting patterns in residue burning as captured by the survey data. This approach is motivated both by the absence of an alternative "ground-truthing benchmark", and because most existing evidence on smallholder farmers' choices and outcomes in the development context relies on survey data, so that finding alternative measures based on remote sensing that, on the one hand, coincide well with the survey data; but, on the other hand, feature much larger coverage at lower cost, provides clear benefits.

### B. *Construction and comparison of different remote-sensing burn indicators*

The construction of crop residue burn indicators from remote-sensing data involves numerous choices and parameter selections, such as thresholds for detection or spectral indices. These decisions often depend on the local context and are made with limited guidance from theory or empirical evidence. Our study starts to address this issue by demonstrating the construction of burn indicators using different approaches, and then testing the performance of these indicators in predicting incidences of residue burning as observed in survey data. For this, we use recently collected, publicly available household survey data from India and compare the recorded incidences of crop residue burning at the beginning of the winter (Rabi) season 2021/22 (i.e., just after the rice harvest) with those implied by the burn indicators constructed from remote-sensing data.

---

[3] For instance, in the World Bank's Living Standards Measurement Study - Integrated Surveys on Agriculture (LSMS-ISA), plot geo-coordinates are manipulated by adding random noise to the latitude and longitude.

[4] In contrast, misreporting may be less of a concern for zero tillage, which is neither illegal nor socially very sensitive.

Both the underlying data sources and methods used in constructing the burn indicators are described in detail below. What should be noted at this point is that our analysis focuses on the roles of two key parameters in the construction of remote-sensing burn indicators. First, we compare indicators constructed using village center coordinates with alternative indicators constructed using additional information on the locations of individual agricultural plots belonging to each village. This generates useful insights, as it is typically unknown from remote-sensing data which plots (or corresponding grid cells) belong to a given village. Thus, when performing analyses based on a combination of geospatial (remote-sensing) data and village or household-level variables, one needs to decide how to aggregate the grid cell data, to merge them with the other variables. Our analysis generates insights that can help guide such decisions.

Second, we leverage our (on-ground) survey data to identify appropriate thresholds to distinguish burned from unburned areas in remote-sensing indicators of crop residue burning, a critical challenge in the construction of such indicators.

### C. Regression analysis of the determinants of crop residue burning

Our investigation of the empirical determinants of crop residue burning in the study region considers two main specifications (as well as several sensitivity checks). First, we estimate plot-level regressions using survey data only; that is, both the dependent variable (burn indicator observed at the plot level) and the regressors (observed either at the plot or household level) are based on farmers' self-reported practices as available in the survey data. The regression model used at this stage can be written as:

$$Burn_{dvij}(survey) = \beta + \gamma\, ZeroTillage_{dvij} + X_{dvi}\delta + W_{dvij}\,\theta + \eta_v + \varepsilon_{dvij}, \qquad (1)$$

where $d$ indexes districts, $v$ villages, $i$ households, and $j$ plots. $Burn_{dvij}$ is an indicator variable that equals one if a household reported to have practiced stubble burning on plot $j$ in the Rabi Season 2021/22 (and zero otherwise). $ZeroTillage_{dvij}$ is defined analogously as an indicator capturing zero tillage on plot $j$. $X_{dvi}$ is a vector of household-level control variables, $W_{dvij}$ is a vector of plot-level controls, and $\eta_v$ denotes village fixed effects. The error term $\varepsilon_{dvij}$ is clustered at the district level (with alternative specifications considered as robustness checks). The main coefficient of interest is $\gamma$, with negative (positive) values of $\gamma$ indicating that the estimated likelihood of burning is smaller (higher) if a plot is cultivated with zero tillage.

An advantage of this approach based solely on survey data is that both the outcome (burn indicator) and key regressor (use of zero tillage) are observed at the plot level, allowing for a precise estimate of the empirical link between the practices of zero tillage and residue burning. At the same time, by using farmers' self-reported burning practices, the validity of this approach depends crucially on the assumptions of perfect recall and truthful reporting in the used survey data. Evidence from previous studies suggests that this assumption is generally not fulfilled in household surveys, and more likely violated when survey items refer to topics with strong social norms or legally binding constraints.

Since the practice of stubble burning is both illegal and socially sensitive in India, relying on survey data for capturing stubble burning may suffers from misreporting biases.[5]

Our second specification therefore uses burn indicators constructed from remote-sensing data (satellite images), which are not subject to these biases (i.e., independent of perfect recall and truthful reporting). However, an important limitation of this approach is that remote-sensing data on residue burning is typically not observed at the plot level, but for larger grid cells which usually comprise multiple plots (or parts thereof). Moreover, the spatial resolution of publicly available indicators is typically not sufficient to capture small fires, and limited temporal resolution (exacerbated by cloud obstructions) may miss to capture short-lived fires.

Even if there were remote-sensing data on burning available of sufficient precision for individual plots (which some studies attempt to construct using visual inspection of satellite images to identify plots), mapping this information to survey data for each plot would require the exact geo-coordinates of each plot. Such plot coordinates are typically unavailable in widely-used agricultural household surveys, either because their collection would be too cumbersome and expensive, or because recorded plot geo-coordinates need to be manipulated by adding random noise to ensure respondents' anonymity, which makes it impossible to identify individual plots and merge them with geocoded variables from other datasets. Thus, given the difficulty in matching remote sensing data to individual plots, the dependent variable in our regression analysis is constructed at the village level, namely by aggregating grid-level data for the plots belonging to each village (as described below). While this approach helps to overcome the above-mentioned misreporting bias in survey data, working with burning data that are aggregated at the village level tends to reduce the precision of the estimates, making it more difficult to find statistically significant effects.[6] Yet, to the extent that we are able to obtain significant estimates, the feature of aggregation is not a first-order concern in identifying the existence (and direction) of the effect of zero tillage on crop residue burning. Formally, the regression model using remote-sensing data is given by:

$$Burn_{dv}(remote\ sensing) = \beta + \gamma\ ZeroTillage_{dvij} + X_{dvi}\delta + W_{dvij}\theta + \psi_d + \varepsilon_{vij}, \qquad (2)$$

where $Burn_{dv}$ is a continuous variable measuring the extent of crop residue burning in village $v$ of district $d$ in the Rabi Season 2021/22 using remote-sensing data. The terms $ZeroTillage_{dvij}$, $X_{dvi}$, and $W_{dvij}$ are observed at the household and plot level, and defined analogously to equation (1). $\psi_d$ denotes district fixed effects, replacing the village fixed effects in equation (1), as the variation in the dependent variable is now limited to the village level (rather than the plot level). The error term $\varepsilon_{vij}$ is once more clustered at the district level. The main coefficient of interest is $\gamma$, with negative (positive) values of $\gamma$ indicating that the extent of burning in the respective village is smaller (higher) if a plot is cultivated using zero tillage.

### III. Data and variables

*A. Household and plot-level survey data*

The survey data for our analysis come from the publicly available dataset described in Naeher et al. (2023), which contains the responses of 1,206 wheat farmers residing in 70 villages in Punjab and

---
[5] This applies less to self-reported data on zero tillage which is neither illegal nor socially very sensitive.
[6] Conceptually, the remote-sensing burn indicator at the village level represents a noisy measure of burning on each plot.

Haryana collected between September and October 2022. The villages had been selected from a sampling frame of 1722 communities in Haryana and Punjab that had been identified as predominantly wheat growing areas based on remote-sensing data from satellite images. For reasons explained below, we restricted the sample to 66 villages (1,143 households) with sufficient data.

Figure 1 depicts the geographic area of our study and the locations of the villages included in the analysis. In each village, up to 28 farm-household observations are available, including rich information about household characteristics and agricultural practices as well as detailed information at the plot level, focusing on the main wheat plot cultivated by each household in the winter (Rabi) season 2021/22 (which lasts from October to June, with sowing usually taking place between late October and November).

Variables observed at the household level include household size (i.e., number of household members), the age, gender, education, religion, and caste of the household head, as well as a binary indicator capturing whether the household owns a tractor or not (which may serve both as a proxy for agricultural technology and for households' wealth more generally). Variables observed at the plot level include plot size (measured in hectares), distance from the plot to the farmer's home (in km), ownership status, tillage method, use of crop residue from the previous season, and binary indicators capturing fertilizer application, early sown wheat (ESW), and hired-in labor, respectively (all referring to the Rabi season 2021/22).

The information on tillage method is provided as a categorical variable with the following response options: (1) cultivator, (2) harrow, (3) rotavator, (4) zero tillage equipment, and (5) other. Based on this, we construct a binary indicator at the plot level for zero tillage, which captures whether zero tillage equipment (response item 4) was used.

Similarly, the data on crop residue usage is available as a categorical variable with the following response options: (1) incorporated residues into the soil, (2) used as soil mulch, (3) burned all residues in the field, (4) burned all residues after heaping together, (5) removed and heaped outside the plot, (6) used as cattle feed, (7) burned only the loose residues, with the rest removed, (8) sold, (9) handed over freely, and (10) other. Based on this, our plot-level burn indicator is constructed as a binary variable that equals one for items (3), (4), and (7), and zero otherwise.

**Figure 1.** Study area and location of sample villages

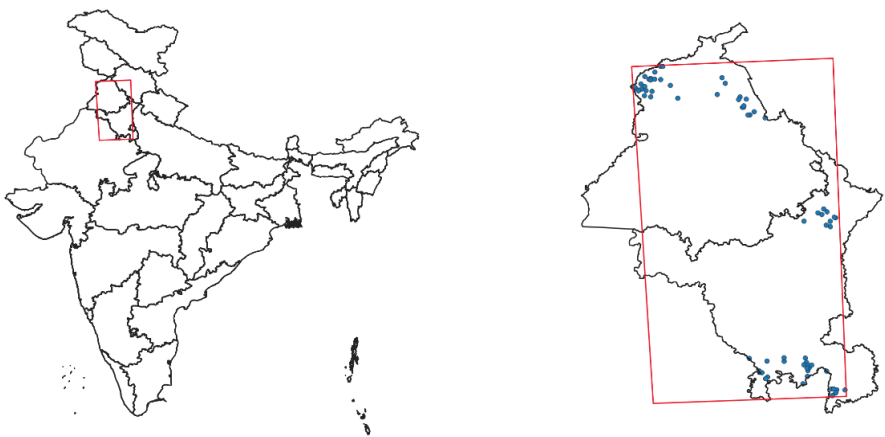

*Notes:* Map of India depicting district boundaries (ADM1) and study area (left). The right panel provides a zoomed-in view of Punjab and Haryana, displaying the study area boundaries and the locations of the villages included in the study.

**Table 1**. Summary statistics of survey variables (household/plot level)

| Variable | Mean | Minimum | Maximum | Std. Dev. | Observations |
|---|---|---|---|---|---|
| HH size | 5.899 | 1 | 38 | 3.689 | 1143 |
| HH head age | 48.551 | 17 | 95 | 13.869 | 1143 |
| HH head male | 0.914 | 0 | 1 | 0.28 | 1143 |
| HH head secondary education | 0.416 | 0 | 1 | 0.493 | 1097 |
| Hindu | 0.471 | 0 | 1 | 0.499 | 1139 |
| Scheduled caste | 0.416 | 0 | 1 | 0.493 | 1133 |
| Tractor | 0.443 | 0 | 1 | 0.497 | 1142 |
| Plot area | 1.618 | 0.031 | 48.48 | 2.324 | 1142 |
| Plot distance | 1.701 | 0 | 80 | 3.938 | 1140 |
| Plot owned | 0.937 | 0 | 1 | 0.244 | 1137 |
| Zero tillage | 0.120 | 0 | 1 | 0.326 | 1138 |
| Residue burning | 0.055 | 0 | 1 | 0.229 | 1119 |
| Plot with ESW | 0.104 | 0 | 1 | 0.305 | 1138 |
| Fertilizer | 0.942 | 0 | 1 | 0.234 | 1135 |
| Outside labor | 0.805 | 0 | 1 | 0.397 | 1136 |

Table 1 provides basic summary statistics for the survey-based variables used in the analysis. The average household in our sample comprises between five to six members and is headed by an adult male. About 44.3% of households own a tractor while 41.6% of household heads have completed secondary education, 47.1% are Hindu, and 41.6% belong to a scheduled caste. The average plot has a size of 1.6 hectares (with median of 0.8 hectares) and is located 1.7 km from the household's home.[7] The use of zero tillage was reported for 12.0% of plots (spread across 47.0% of villages in the sample). Residue burning was reported for 5.5% of plots (spread across 40.9% of villages). Most plots are owned by the household (93.7%) and were cultivated using fertilizer (94.2%) and hired-in labor (80.5%).

The household survey dataset also includes the GPS coordinates (latitude and longitude) of 515 plots,[8] which we use in mapping the survey data to the remote-sensing data as described below. Since the burn indicators based on remote-sensing data will be constructed at the village level, we also create a survey-based burn indicator at the village level, which captures, for each village, the share of plots in the sample that were subject to residue burning.

### B. *Geospatial and remote-sensing data*

***Satellite imagery.*** The remote sensing data for this analysis were obtained from Sentinel-2, a satellite mission developed by the European Space Agency (ESA) as part of the Copernicus program. Sentinel-2 imagery is publicly accessible, providing a valuable resource for a wide range of research applications. The high spatial resolution of the imagery—10 m for visible and near infrared (NIR) bands, and 20 m for red-edge and shortwave infrared (SWIR) bands—enables effective detection of short-lived and spatially localized fire events.

For this study, the data were sourced from the Sentinel-2 Harmonized Surface Reflectance dataset (Claverie et al., 2018), available through the Google Earth Engine (GEE) platform (Gorelick et al., 2017). This dataset provides Level-2A products, which are atmospherically corrected to ensure consistent surface reflectance values, enhancing the reliability of environmental analyses.

---
[7] One plot area with an unusually large size (1355.4 hectare) was excluded from the analysis.
[8] Missing GPS coordinates for some plots were due to inaccessibility (e.g., flooded roads), time constraints during the survey, and lack of consent by the farmer.

To focus the analysis on relevant areas, a bounding box was defined to encompass all villages and plots within the study area. Figure 1 depicts the location of our study area. Sentinel-2 images were extracted for the period from September to November 2021, yielding 394 tiles after filtering for cloud cover (limited to 20% or less). Several preprocessing steps were applied to ensure the reliability of the dataset. First, cloud and cirrus contamination were removed using the 'QA60' band, retaining only clear-sky observations. Water bodies were then masked out using the 'max-extent' band of the JRC Global Surface Water dataset (Pekel et al., 2016). This preprocessing strategy effectively mitigates interference from clouds and water, ensuring a robust dataset for analyzing fire occurrence and spatial patterns in the study area (Roy et al., 2019; Deshpande et al., 2022).

Finally, since this analysis focuses on agricultural areas, urban regions were excluded using a binary urban classification mask derived from the Global Human Settlement Layer (GHSL) dataset (Pesaresi et al., 2023). The GHSL dataset measures the "built-up fraction" (BUFRAC), which represents the proportion of a raster cell covered by built-up structures. As specified in the dataset's documentation, any BUFRAC value exceeding 50 is classified as urban. For this analysis, the 2018 dataset, which provides a 10m spatial resolution consistent with Sentinel-2 data, was used. While the GHSL dataset is available for more recent years, these versions offer a coarser spatial resolution. Using the BUFRAC value, a binary mask was created to classify areas as urban (1) or rural (0). To ensure the mask's accuracy, it was validated through a visual comparison with high-resolution satellite imagery from Google Earth, revealing a strong correlation and confirming its effectiveness in excluding urban areas from the analysis.

*MODIS Burned Area data.* To calibrate the thresholds of our burn indicators, we used the Terra and Aqua combined MCD64A1 Version 6.1 Burned Area product (Giglio et al., 2018), a monthly global dataset at 500m resolution. This product provides per-pixel burned-area and quality information based on MODIS Surface Reflectance imagery and active fire observations. This dataset allowed us to observe that, for our study area and period, burned pixels were detected only during weeks 35 (September 1–7), 39 (September 29–October 5), and 43 (October 27–November 2).

*VIIRS NPP Active fire data.* To further validate our burn indicators, we utilized the VIIRS NPP Active Fire dataset (Giglio, 2024; Schroeder et al., 2014), which is a key component of NASA's Suomi-NPP VIIRS Active Fire product suite. This dataset builds upon the MODIS Fire and Thermal Anomalies algorithm (MOD14/MYD14) and provides global fire activity data at intervals of 12 hours or less. Among the available resolutions—750 m and 375 m—we selected the finer 375 m product for its enhanced spatial detail. Although the 375 x 375 m resolution is too coarse for precise burn area calculations within our study areas, it was invaluable for calibration and validation.

The active fire data was downloaded in shapefile format for the period from September 1 to November 30, 2021. After clipping the data to our area of interest, 77,664 points representing active fires remained within the study area. To ensure data reliability, we retained only high-confidence detections (labeled as CONFIDENCE = 'h') and focused on agricultural fires (TYPE = 0). This filtering reduced the dataset to 3,541 high-confidence observations.

A key step in data preparation was the exclusion of points located in urban areas, as detections in these regions may not reflect agricultural burning. To accomplish this, we applied the binary urban mask described earlier, classifying each detection as either urban or rural based on its geographic location. Only points classified as rural were retained, resulting in 3,519 points.

Each of these points represents the center of a pixel flagged for containing a fire or other thermal anomaly. In other words, the "location" refers to the center of the pixel, rather than the exact coordinates of the fire (Earth Science Data Systems, 2024). To approximate the area where the fire was detected, square polygons of 375 x 375 m (matching the spatial resolution of the active fire data) were created around each rural detection point.

**IV.     Construction of remote-sensing burn indicators**

The use of remote sensing-derived indicators for quantifying agricultural burns is a well-established practice (see the references cited in the Introduction). Agricultural fires can generally be detected from space either by measuring radiation in the thermal bands (Wooster et al., 2013) or by observing the aftermath of fires, such as reduced vegetation cover or charred surfaces. To support fire detection, various spectral indices have been developed using remote sensing data. Among the most widely applied are the Normalized Burn Ratio (NBR) (Roy et al., 2006), the Burn Area Index (BAI) (Chuvieco et al., 2002; Liu et al., 2021), and the Sentinel-2 Burn Area Index (BAIS2) (Filipponi, 2018). Each index has its strengths and limitations. For instance, the Normalized Burn Ratio (NBR) is well-suited for detecting burned areas shortly after a fire event but tends to lose effectiveness as vegetation regrows. In contrast, the Burn Area Index (BAI) is more sensitive to the char signal in post-fire imagery, making it better at distinguishing burned areas from other land cover types (Liu et al. 2021). BAIS2 was developed to leverage the spectral properties of the Sentinel-2 MSI sensor, using a tailored combination of bands demonstrated to perform well in detecting burned areas in post-fire landscapes (Filipponi 2018).

In developing our remote-sensing burn indicators, we evaluated all three approaches and tested their performance for our analysis. The full procedure is documented in Appendix A. As detailed there, BAIS2 emerged as the most effective index for distinguishing burned from unburned areas in our study region. To further validate the results, we applied the defined thresholds to the Sentinel-2 tiles and compared the outcomes with VIIRS active fire data for the three weeks during which MODIS data was available. As illustrated in Figure 2, there is a strong overlap between the thresholded burned areas and the VIIRS detections.

**Figure 2**. Validation of Burnt Mask

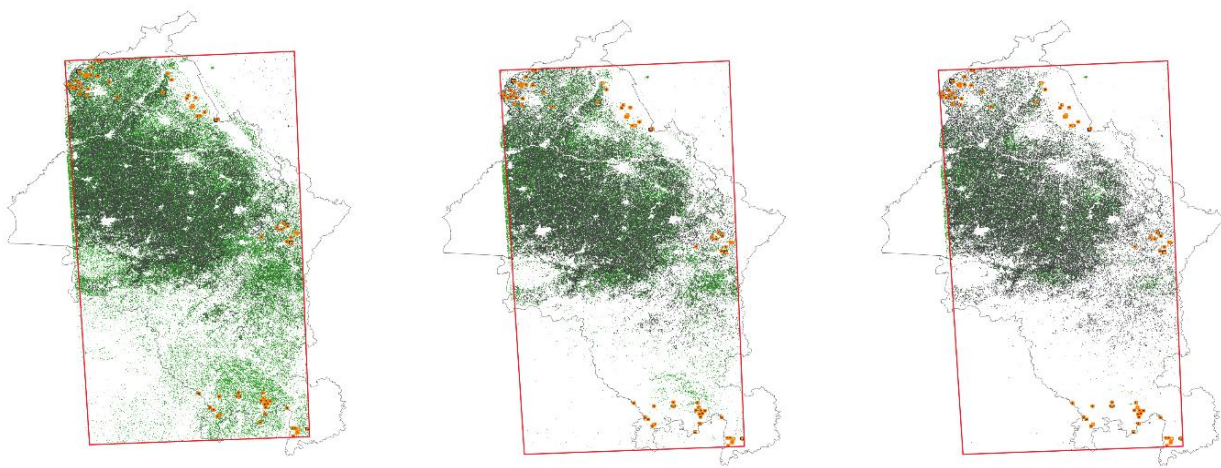

*Notes*: The map shows in green areas identified as burned based on our thresholding method (left: threshold = 0.85, middle = threshold = 0.90, right = threshold = 0.95), overlaid with VIIRS active fire data (in gray), and the administrative boundaries of Punjab and Haryana. Villages included in the analysis, based on both village-level and plot-level methods, are indicated in orange and red, respectively.

## V. Results

### A. *Comparison of remote-sensing and survey-based burn indicators*

Table 2 reports pairwise correlations between the survey-based burn indicator and the six remote-sensing burn indicators. The highest correlation with the survey-based indicator is obtained under a BAIS2 threshold of 0.90, both among the three indicators constructed using village coordinates, and among the three indicators based on plot coordinates. Our subsequent analysis thus focuses on these two indicators, while the indicators with thresholds 0.85 and 0.95 will be used for sensitivity checks.

**Table 2**. Pairwise correlations between burn indicators

|  | Survey data | Village coord. (BAIS2>0.85) | Village coord. (BAIS2>0.90) | Village coord. (BAIS2>0.95) | Plot coord. (BAIS2>0.85) | Plot coord. (BAIS2>0.90) |
|---|---|---|---|---|---|---|
| Village coord. (BAIS2>0.85) | 0.2467 | 1.000 | | | | |
| Village coord. (BAIS2>0.90) | 0.4514 | 0.7661 | 1.000 | | | |
| Village coord. (BAIS2>0.95) | 0.3564 | 0.5108 | 0.8434 | 1.000 | | |
| Plot coord. (BAIS2>0.85) | 0.2892 | 0.7493 | 0.5377 | 0.2718 | 1.000 | |
| Plot coord. (BAIS2>0.90) | 0.4224 | 0.6467 | 0.8276 | 0.6160 | 0.7193 | 1.000 |
| Plot coord. (BAIS2>0.95) | 0.3377 | 0.4262 | 0.7118 | 0.7512 | 0.3913 | 0.7699 |

*Notes*: All burn indicators are constructed at the village level. The survey-based indicator aggregates information reported for individual plots. In all cases, NBR is thresholded to preserve only values which are less than 0.20.

Figure 3 shows the distributions of the remote-sensing burn indicators (BAIS2>0.90 and NBR<0.20) based on village and plot coordinates, respectively. The overall pattern is similar and in line with the strong correlation (0.83) reported in see Table 2. This is also visible in Figure 4, which shows a scatter plot of the relationship between the two indicators. While the village-based indicator is always strictly greater than zero (i.e., there is at least one grid cell with some amount of burning in each village), the plot-based indicator equals zero for three village observations.

In the survey data, 27 of 66 villages feature at least one plot with residue burning. Of these, 26 villages (96.3%) also show burning according to the remote-sensing indicator based on plot coordinates (BAIS2>0.90 and NBR<0.20). Moreover, of the 39 villages without burning in the survey data, two villages (5.1%) also feature no burning in the remote-sensing indicator, while 37 villages (94.9%) do show burning. This discrepancy is to be expected, given that the remote-sensing indicator captures the complete area around each village, whereas the survey data captures burning only for a small subset of plots belonging to each village. Overall, these results thus suggest potentially large benefits of using remote-sensing indicators to detect crop residue burning, as (a) these coincide well with the instances of burning captured in the survey data while at the same time allowing for detecting instances of burning on other plots than those captured in the survey data, and (b) offering much larger geographical and temporal coverage (and typically at lower cost) than household surveys. Notice that (a) generally applies for all on-ground data which are collected only for a subset of plots belonging to each village (a typical feature of agricultural surveys).

**Figure 3**. Distribution of remote-sensing burn indicators using village coordinates (left) and plot coordinates (right)

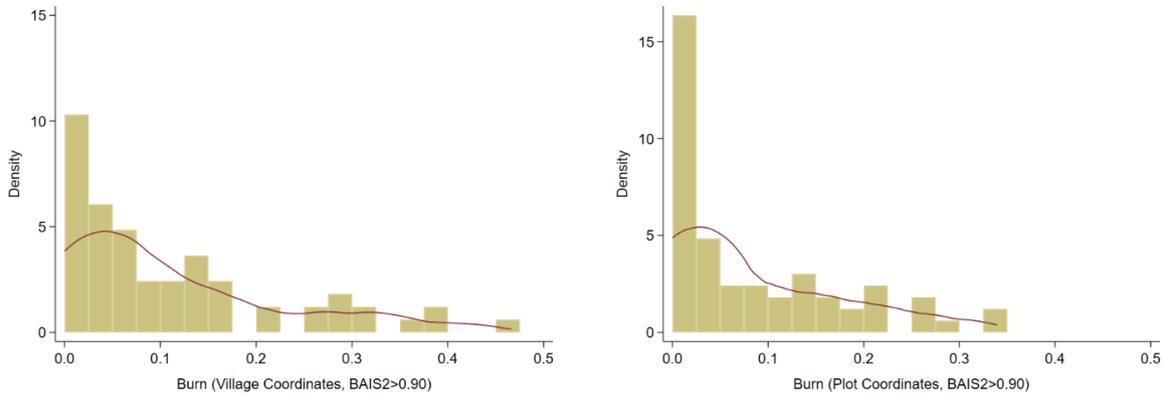

*Notes*: The red line in each chart depicts a Kernel density estimate.

**Figure 4**. Scatter plot of the relation between village-based and plot-based residue burn indicators

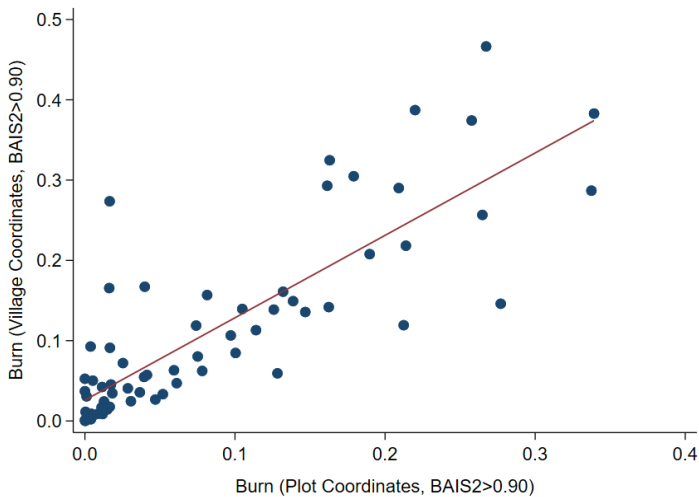

*Notes*: The red line depicts the fitted values of a bivariate linear regression.

### B. *Plot-level regression analysis using survey data*

We now turn to investigating the empirical relationship between zero tillage and crop residue burning, considering first the plot-level regression model using survey data only, and then the specification that uses burn indicators constructed from remote-sensing data.

Table 3 reports the results obtained from estimating the plot-level regression model based on survey data specified in equation (1). Since the dependent variable in this specification is binary and the estimation is performed using ordinary least squares, the reported coefficients can be interpreted as a linear probability model. Column (1) of Table 3 shows the bivariate regression of crop residue burning on zero tillage use. The estimated coefficient is negative and statistically highly significant, indicating that the practice of stubble burning is less prevalent on plots that are cultivated with zero tillage. This result remains intact when household characteristics are added as controls to the regression, including demographic factors (column 2) and a dummy capturing ownership of a tractor (column 3). Our preferred specification further includes various plot control variables as reported in column (4). The coefficient on zero tillage estimated from this specification indicates that plots that are cultivated with zero tillage are 3.9 percentage points less likely to be subject to crop residue burning. Given that the baseline share of plots with residue burning is 5.5% (see Table 1), this amounts to a reduction in the likelihood of burning greater than two thirds (70.9%), making zero tillage a very strong predictor of stubble burning.

In addition, the results in column (4) of Table 3 suggest that stubble burning is less likely to occur on plots that are owned by the household (as opposed to leased or shared in). Moreover, the fact that none of the included household characteristics is statistically significantly associated with residue burning suggests that the practice of stubble burning in this region of India is not limited to certain groups of farmers (e.g., based on education, caste, or religious background), but seems to occur largely independent of these factors.

**Table 3.** Results for plot-level regressions using survey data – Dependent variable: Residue burning (binary indicator at the plot level, survey data)

|  | (1) | (2) | (3) | (4) |
|---|---|---|---|---|
| Zero tillage | -0.030*** | -0.044*** | -0.044*** | -0.039*** |
|  | (0.000) | (0.007) | (0.009) | (0.002) |
| HH size |  | 0.001 | 0.001 | 0.001 |
|  |  | (0.726) | (0.607) | (0.638) |
| HH head male |  | 0.013 | 0.014 | 0.016 |
|  |  | (0.701) | (0.677) | (0.611) |
| HH head age |  | -0.000 | -0.000 | -0.000 |
|  |  | (0.688) | (0.665) | (0.712) |
| HH head secondary education |  | -0.000 | -0.000 | -0.001 |
|  |  | (0.989) | (0.994) | (0.934) |
| Hindu |  | 0.021 | 0.021 | 0.024 |
|  |  | (0.177) | (0.192) | (0.145) |
| Scheduled caste |  | 0.008 | 0.008 | 0.002 |
|  |  | (0.891) | (0.896) | (0.976) |
| Tractor |  |  | -0.010 | -0.008 |
|  |  |  | (0.587) | (0.624) |
| Plot area |  |  |  | -0.002 |
|  |  |  |  | (0.682) |
| Plot distance |  |  |  | -0.000 |
|  |  |  |  | (0.632) |
| Plot with ESW |  |  |  | -0.020 |
|  |  |  |  | (0.341) |
| Plot owned |  |  |  | -0.049** |
|  |  |  |  | (0.020) |
| Fertilizer |  |  |  | 0.002 |
|  |  |  |  | (0.931) |
| Outside labor |  |  |  | 0.005 |
|  |  |  |  | (0.841) |
| Village FE | yes | yes | yes | yes |
| Observations | 1,114 | 1,059 | 1,059 | 1,053 |
| R-squared | 0.200 | 0.220 | 0.220 | 0.226 |
| R-squared adj. | 0.149 | 0.163 | 0.163 | 0.163 |
| Mean(dependent variable) | 0.055 | 0.055 | 0.055 | 0.055 |

Note: $p$-values in parentheses: * $p < 0.10$, ** $p < 0.05$, *** $p < 0.01$. Estimated using ordinary least squares. Standard errors are clustered at the district level.

*C. Village-plot level regressions using remote-sensing and survey data*

Table 4 reports the results obtained from estimating the regression model specified in equation (2), which uses both survey data (observed at the household and plot level) and the remote-sensing burn indicator constructed at the village level (using the geo-coordinates of the plots belonging to each village). Notice that, differing from the regressions in the previous subsection, the dependent variable now is a *continuous* measure of crop residue burning (as described in Section V). Other than that, the results reported in columns (1) to (4) of Table 4 follow the same logic as before, starting with the bivariate regression of crop residue burning on zero tillage, and then subsequently adding further controls. The estimated coefficient on zero tillage is always negative and statistically significant at the 5% significance level, confirming the finding from above that stubble burning happens significantly less on plots that are cultivated with zero tillage. More specifically, the coefficient on zero tillage reported in column (4) indicates that plots that are cultivated with zero tillage feature, on average, a 0.39 smaller value of the remote-sensing burn indicator than plots without zero tillage. Given that the baseline value of this burn indicator is 0.078, this amounts to a reduction in the extent of stubble burning of one half (50.0%).

    The results in column (4) of Table 4 also confirm the finding that the ownership status of plots matters, with stubble burning being less likely to occur on plots that are owned by the household. In addition, there is some evidence that the incidence of stubble burning increases for households belonging to a scheduled caste as well as for plots with ESW (both statistically significant at the 10% significance level). The latter finding is interesting, as it appears to support the common concern in the literature that ESW may necessitate stubble burning as a means of saving time in land preparation to facilitate early sowing. However, as mentioned above and in line with the empirical evidence we obtain, the use of zero tillage equipment (such as Happy seeder, Super seeder, Turbo seeder, etc.) makes the step of stubble burning obsolete, including when wheat is sown early. Overall, our results may thus rather be interpreted as highlighting the complementarity between zero tillage and ESW in moving towards environmentally friendly wheat farming, including by reducing the incidence of crop residue burning.

    *Sensitivity*. The finding of a negative association between zero tillage and crop residue burning continues to hold when using the burn indicator with a BAIS2 threshold of 0.95, but ceases to be statistically significant for a BAIS2 threshold of 0.85 (see Tables B1 and B2 in Appendix B). Moreover, none of the burn indicators constructed based on village coordinates yields statistically significant results (not shown). This highlights that the parameters and choices underlying the construction of the remote-sensing based burn indicators are crucial.

**Table 4.** Results for village-plot-level regressions – Dependent variable: Residue burning (continuous indicator at the village level, remote-sensing using plot coordinates, BAIS2>0.90)

|  | (1) | (2) | (3) | (4) |
|---|---|---|---|---|
| Zero tillage | -0.043** | -0.042** | -0.042** | -0.039** |
|  | (0.023) | (0.015) | (0.015) | (0.022) |
| HH size |  | -0.001 | -0.001 | -0.001 |
|  |  | (0.436) | (0.401) | (0.408) |
| HH head male |  | 0.005 | 0.005 | 0.005 |
|  |  | (0.536) | (0.530) | (0.559) |
| HH head age |  | 0.000 | 0.000 | 0.000 |
|  |  | (0.183) | (0.176) | (0.159) |
| HH head secondary education |  | 0.003 | 0.003 | 0.002 |
|  |  | (0.398) | (0.406) | (0.576) |
| Hindu |  | 0.008 | 0.008 | 0.007 |
|  |  | (0.410) | (0.422) | (0.473) |
| Scheduled caste |  | 0.006 | 0.006* | 0.005* |
|  |  | (0.106) | (0.093) | (0.097) |
| Tractor |  |  | -0.000 | 0.000 |
|  |  |  | (0.970) | (0.911) |
| Plot area |  |  |  | -0.000 |
|  |  |  |  | (0.684) |
| Plot distance |  |  |  | -0.000 |
|  |  |  |  | (0.840) |
| Plot with ESW |  |  |  | 0.017* |
|  |  |  |  | (0.054) |
| Plot owned |  |  |  | -0.015* |
|  |  |  |  | (0.097) |
| Fertilizer |  |  |  | -0.003 |
|  |  |  |  | (0.803) |
| Outside labor |  |  |  | 0.002 |
|  |  |  |  | (0.722) |
| Village FE | yes | yes | yes | yes |
| Observations | 1,143 | 1,086 | 1,086 | 1,075 |
| R-squared | 0.505 | 0.511 | 0.511 | 0.513 |
| R-squared adj. | 0.503 | 0.505 | 0.505 | 0.504 |
| Mean(dependent variable) | 0.078 | 0.078 | 0.078 | 0.078 |

*Note: p-values in parentheses: * $p < 0.10$, ** $p < 0.05$, *** $p < 0.01$. Estimated using ordinary least squares. Standard errors are clustered at the district level.*

## VI. Conclusion

This study introduces a novel method for enhancing the construction of remote-sensing indicators for crop residue burning at the plot level. Our approach combines high-resolution, publicly available satellite imagery with precise household and plot-level spatial data from survey-based data collection. This integration allows us to empirically assess the relationship between crop residue burning and the adoption of zero tillage practices.

Applying the proposed approach to our case study of 1,143 households in Punjab and Haryana (Northwest India) provides valuable insights that underscore both the effectiveness and replicability of the method. First, we confirm that remote-sensing burn indicators exhibit strong correlations with farmers' self-reported data on crop residue burning. This is a key finding, given that survey data is not always readily available, while remote-sensing data is generally more accessible, including at higher spatial and temporal resolutions. Additionally, our analysis demonstrates that incorporating plot-level coordinates significantly improves the accuracy of remote-sensing burn indicators. This is particularly important as it enhances the precision of burn pattern predictions, making the method more reliable. We also establish precise, thoroughly validated thresholding ratios for spectral indices, which can be immediately implemented in similar studies, saving both time and computational resources by eliminating the need for testing and validation of various indices and parameters.

Finally, our results reveal a robust, negative, and statistically significant relationship between the adoption of zero tillage and reduced crop residue burning, solidifying a common perception in the literature that zero tillage reduces the incidence of stubble burning, as observed both in farmers' self-reports and remote-sensing data. When focusing on a survey-based measure of residue burning, our estimates indicate that plots that are cultivated with zero tillage are 3.9 percentage points less likely to be subject to crop residue burning – a large reduction compared to the baseline share of 5.5% of plots with residue burning. When using a remote-sensing burn indicator, we find that zero tillage is associated with a reduction in the extent of stubble burning of one half.

We stress again that the analysis is subject to several limitations, including challenges related to the availability and quality of both geolocation data and high-resolution satellite imagery (e.g. due to cloud cover in some regions). Furthermore, potential spatial and temporal mismatches between the survey data and remote sensing data pose challenges to ensuring precise alignment, potentially affecting the accuracy of the derived burn indicators. These limitations underscore the need for more systematic data collection practices and robust data validation. Moreover, our analysis highlights the sensitivity of the results to the choice of parameters, such as the thresholds used for remote-sensing burn indicators, emphasizing the importance of careful parameter selection. Furthermore, while our study provides valuable insights into the specific context of Punjab and Haryana, these findings may not be directly transferable to other regions without careful adaptation to local geographic, ecological, and socio-economic conditions.

More research is clearly warranted to address these limitations and explore the broader applicability of our methodology. A specific consideration for further research is the reduced effectiveness of the village-centered indicator in comparison to the indicator derived from plot coordinates. We hypothesize that the reduced effectiveness of the village-level indicator is due to the aggregation of data from plots to villages (villages often contain a mix of plots with and without zero tillage), which smooths out variations in the relationship between zero tillage practices and residue burning. Given that village-level data is easier to collect than plot-level data, this presents a critical area for future work to refine methods for leveraging village-level data. More generally, one promising direction is to expand the scope of analysis to include a wider range of geographic, ecological, and agricultural contexts. This would help validate the robustness and generalizability of our approach, particularly in regions with different agricultural practices, climates, and data availability. Additionally, future work could explore the integration of more diverse datasets, such as socio-economic and environmental data, to further enhance the accuracy and predictive power of remote-sensing indicators.

**Appendix A: Construction of remote-sensing burn indicators**

*A.1  Polygons for the areas of interest*

To derive indicators for each unit of analysis, we constructed polygons for the villages using both the village-level and plot-plot data included in the study. Both types of polygons were validated by overlaying them with high-resolution satellite imagery from Google Earth to assess their accuracy.

*Village-level polygons.* The survey data included 1,206 observations corresponding to 70 unique villages, each associated with latitude and longitude coordinates. These coordinates represent the center of each village, calculated as the average of the plot coordinates within that village. To define the analysis area, bounding boxes were created around these village centers, with each box having a square shape and equal area. The size of the bounding boxes was determined by calculating the distances between the village center and the plot coordinates associated with the same village.

During this process, three observations with unusually large distances (~80 km) were identified as likely data errors or special outlier cases and excluded from the calculations. After removing these outliers, the maximum distance between a plot and its village center was approximately 4,600 m, with an average distance of around 1,200 m. Initial bounding boxes were constructed using these distances as radii, but the maximum distance resulted in overly large areas with significant overlap between regions of interest. Conversely, using the average distance reduced the overlap but excluded some plot coordinates.

To balance coverage and overlap, two additional bounding boxes were created, incorporating a 10% and 20% margin based on the average distance. After evaluating these options, the bounding box based on the average distance with a 20% margin was selected. This approach, illustrated in Figure A1, yielded 70 bounding boxes, each covering an area of approximately 8 km², providing a consistent framework for analysis.

*Plot-based polygons.* The plot-based polygons were constructed using latitude and longitude coordinates associated with individual plots within each village. Of the 70 villages included in the analysis, plots from 3 villages lacked coordinate data and were therefore excluded. Additionally, one more village was excluded due to the large distances between its plot coordinates and the village center, as previously explained. This resulted in 66 areas (villages) of analysis.

To define these areas, bounding boxes were created as the smallest rectangular regions that completely enclosed the plot coordinates for each village. A 20% margin, based on the average area of these bounding boxes, was then added to provide additional coverage. In three cases, the bounding boxes were too small to be effectively analyzed with remote sensing data. To address this, these boxes were proportionally increased to ensure a minimum area of 150,000 m², aligning with the spatial resolution of the satellite imagery used for validation (375x375 m).

The resulting plot-based polygons, illustrated in Figure A1, varied in size and did not overlap with each other. In general, they were smaller than the polygons derived from village-level coordinates, except in three instances where the plot-based areas were larger. This variation ensured that the analysis areas were appropriately scaled to capture plot-level details while remaining compatible with the available remote sensing data.

**Figure A1.** Comparison of Village-Based and Plot-Based Analysis Polygons

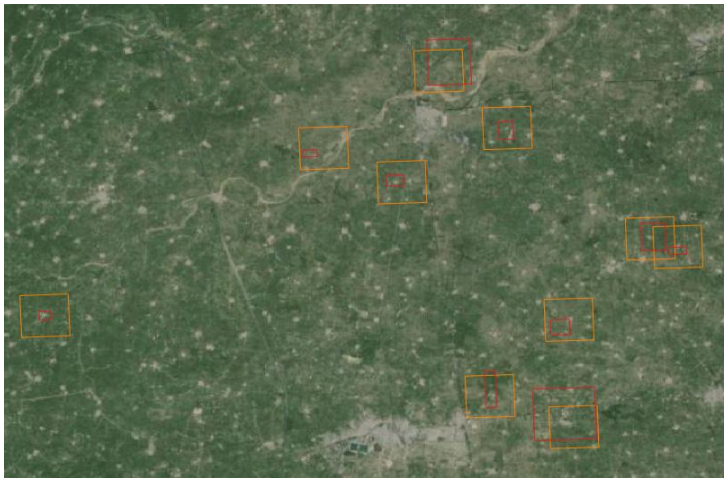

*Notes:* This figure illustrates the spatial differences between the two approaches used to define analysis areas: village-based polygons (shown in orange) and plot-based polygons (shown in red). A satellite image from Google Earth serves as the background for visual reference. The figure highlights differences in shape, size, and alignment between the two approaches.

*A.2  Remote-sensing burn indicators*

Various spectral indices using remote sensing data have been developed in the literature. Among the most widely applied are the Normalized Burn Ratio (NBR), the Burn Area Index (BAI), and the Sentinel-2 Burn Area Index (BAIS2) as defined in Table A1. We calculated these for all available satellite images within the study area during the specified period, encompassing 364 tiles. Each index was added as an additional band to the corresponding tile. To better understand the distribution and behavior of each index across space and time, we computed summary statistics including minimum, maximum, mean, and median values. These metrics provided initial insights into the range, central tendency, and variability of each index, which informed our assessment of their effectiveness for burned area detection.

**Table A1**. Commonly used spectral indices to detect fires from optical satellite imagery

| Index | Acronym | Formula | Primary Use |
|---|---|---|---|
| **Normalized Burn Ratio** | NBR | NBR = (NIR − SWIR) / (NIR + SWIR) | Detecting burned areas and fire severity by measuring vegetation and soil moisture changes. |
| **Burned Area Index** | BAI | BAI = 1 / [(RED − 0.1)$^2$ + (NIR − 0.06)$^2$] | Identifying burned areas by measuring spectral contrast with vegetation. |
| **Burned Area Index for Sentinel-2** | BAIS2 | BAIS2 = (SWIR2 − RED) / (SWIR2 + RED) | Optimized for Sentinel-2 data to detect burned areas using shortwave infrared and red bands. |

*Notes*: NIR = near infrared, SWIR = shortwave infrared.

**Figure A2**. Example of true-color (left) and false-color (right) composites of a section within the study area, generated using median values from October 2021.

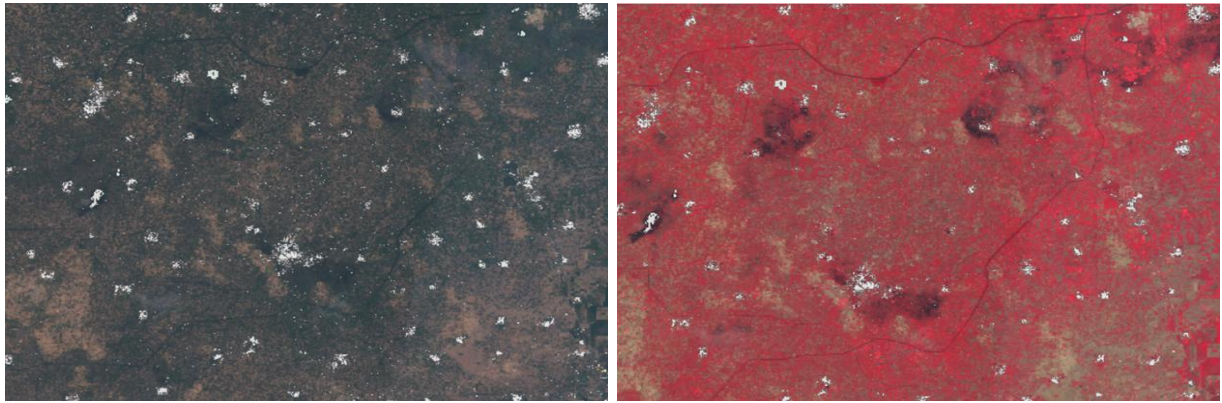

***Sampling strategy***. Identifying appropriate thresholds to distinguish burned from unburned areas is a critical challenge when using these indices. While thresholds suggested in the literature (e.g., Filipponi, 2018) provide a useful starting point, they are often context-specific, reflecting the geographic and seasonal conditions of particular study areas.

To address this limitation and ensure the thresholds were valid for our study area, we adopted a sampling strategy inspired by Deshpande et al. (2022). To align the temporal and spatial resolution of our data sources and streamline processing, we generated weekly median composites for both MODIS and Sentinel-2 data. These composites focused on the three weeks during which MODIS detected burned areas. Sentinel-2 composites were created in both true color (using the B4 [red], B3 [green], and B2 [blue] bands) and false color (using the B8 [NIR], B4 [red], and B3 [green] bands). False color composites are particularly effective for fire analysis as they emphasize burned areas by leveraging spectral differences in how burned vegetation reflects light, making fire scars more visible and easier to analyze compared to true color composites. Figure A2 illustrates the differences between true and false color composites.

To validate and refine burned area detection, we overlaid the weekly Sentinel-2 and MODIS composites. This visual comparison allowed us to align fire activity detected by MODIS with burn scars visible in Sentinel-2 data. Burned and unburned pixels from Sentinel-2 data were sampled based on three criteria: (1) the pixel overlapped spatially with a MODIS-detected burned pixel during the same time window, (2) it appeared dark brown or black in the false-color composite, and (3) visual inspection of the true-color composite corroborated the interpretation.

***Thresholding***. Based on our analysis and consistent with our previous observations, BAIS2 emerged as the most effective index for distinguishing between burned and unburned areas. While NBR also performed reasonably well, it occasionally misclassified burned areas as uncultivated fields. Building on these findings, we established an optimal BAIS2 threshold of > 0.90, with NBR < 0.20 serving as a secondary criterion to exclude outliers where both indices were simultaneously high or low. Notably, the BAIS2 threshold we identified is similar to the value proposed by Filipponi (2018), who found a threshold of 0.865 for identifying wildfires in the Sicily region of southern Italy during July 2017. For the purpose of conducting sensitivity analysis, we also calculated indicators with BAIS2 thresholds of 0.85 and 0.95 while keeping the NBR value constant.

***Final Validation***. To further validate our results, we applied the defined thresholds to the Sentinel-2 tiles and compared the outcomes with VIIRS active fire data for the three weeks during which MODIS data was available. The analysis revealed a strong overlap between the thresholded burned areas and the VIIRS data. Figure 2 illustrates this comparison. It should be noted that it is expected that the thresholded Sentinel-2 images, with their higher spatial resolution of 10 m compared to VIIRS's 375 m resolution, are expected to provide a more granular detection of fires.

***Zonal statistics***. To identify burned areas across the entire study period and area, we applied the defined thresholds to all Sentinel-2 tiles, creating a binary mask where pixels meeting the criteria were classified as burned. This process was applied to weekly composite images spanning weeks 35 through 47 (our complete study period). The resulting binary layers from each week were then combined into a maximum composite, ensuring that any burned areas detected during the study period were retained. Using this final composite, we applied zonal statistics to calculate a continuous burn indicator, capturing the percentage of burn area for each village. This procedure was performed using (a) the village-level and (b) the plot-level polygons, as well as the three different BAIS2 thresholds described earlier, yielding an overall set of six remote-sensing burn indicators.

**Appendix B: Additional results**

**Table B1**. Results for village-plot-level regressions – Dependent variable: Residue burning (continuous indicator at the village level, remote-sensing using plot coordinates, BAIS2>0.85)

|  | (1) | (2) | (3) | (4) |
|---|---|---|---|---|
| Zero tillage | -0.071 | -0.062 | -0.060 | -0.055 |
|  | (0.300) | (0.344) | (0.334) | (0.389) |
| HH size |  | 0.001 | 0.001 | 0.001 |
|  |  | (0.581) | (0.689) | (0.697) |
| HH head male |  | 0.003 | 0.001 | 0.001 |
|  |  | (0.833) | (0.914) | (0.932) |
| HH head age |  | -0.000 | -0.000 | -0.000 |
|  |  | (0.867) | (0.917) | (0.980) |
| HH head secondary education |  | -0.007 | -0.007 | -0.009 |
|  |  | (0.434) | (0.444) | (0.288) |
| Hindu |  | 0.017 | 0.018 | 0.019 |
|  |  | (0.783) | (0.763) | (0.754) |
| Scheduled caste |  | 0.003 | 0.004 | 0.001 |
|  |  | (0.832) | (0.786) | (0.950) |
| Tractor |  |  | 0.011 | 0.009 |
|  |  |  | (0.160) | (0.231) |
| Plot area |  |  |  | 0.001 |
|  |  |  |  | (0.145) |
| Plot distance |  |  |  | -0.001 |
|  |  |  |  | (0.229) |
| Plot with ESW |  |  |  | 0.006 |
|  |  |  |  | (0.540) |
| Plot owned |  |  |  | -0.041 |
|  |  |  |  | (0.149) |
| Fertilizer |  |  |  | 0.016 |
|  |  |  |  | (0.580) |
| Outside labor |  |  |  | 0.011 |
|  |  |  |  | (0.333) |
| Village FE | yes | yes | yes | yes |
| Observations | 1,143 | 1,086 | 1,086 | 1,075 |
| R-squared | 0.645 | 0.641 | 0.642 | 0.646 |
| R-squared adj. | 0.643 | 0.637 | 0.638 | 0.640 |
| Mean(dependent variable) | 0.078 | 0.078 | 0.078 | 0.078 |

*Note:* p-values in parentheses: $^*\ p < 0.10$, $^{**}\ p < 0.05$, $^{***}\ p < 0.01$. Estimated using ordinary least squares. Standard errors are clustered at the district level.

**Table B2**. Results for village-plot-level regressions – Dependent variable: Residue burning (continuous indicator at the village level, remote-sensing using plot coordinates, BAIS2>0.95)

|  | (1) | (2) | (3) | (4) |
|---|---|---|---|---|
| Zero tillage | -0.034** | -0.032** | -0.033** | -0.033*** |
|  | (0.022) | (0.013) | (0.011) | (0.009) |
| HH size |  | -0.000 | 0.000 | 0.000 |
|  |  | (0.764) | (0.780) | (0.992) |
| HH head male |  | 0.001 | 0.001 | 0.001 |
|  |  | (0.640) | (0.515) | (0.417) |
| HH head age |  | -0.000 | -0.000 | -0.000 |
|  |  | (0.299) | (0.228) | (0.170) |
| HH head secondary education |  | 0.001 | 0.001 | 0.001 |
|  |  | (0.244) | (0.314) | (0.352) |
| Hindu |  | 0.003 | 0.002 | 0.002 |
|  |  | (0.213) | (0.265) | (0.297) |
| Scheduled caste |  | 0.001 | 0.001 | 0.001 |
|  |  | (0.333) | (0.352) | (0.320) |
| Tractor |  |  | -0.002 | -0.002 |
|  |  |  | (0.235) | (0.239) |
| Plot area |  |  |  | -0.000 |
|  |  |  |  | (0.290) |
| Plot distance |  |  |  | -0.000 |
|  |  |  |  | (0.196) |
| Plot with ESW |  |  |  | 0.003 |
|  |  |  |  | (0.148) |
| Plot owned |  |  |  | 0.002 |
|  |  |  |  | (0.320) |
| Fertilizer |  |  |  | -0.002 |
|  |  |  |  | (0.340) |
| Outside labor |  |  |  | -0.001 |
|  |  |  |  | (0.239) |
| Village FE | yes | yes | yes | yes |
| Observations | 1,143 | 1,086 | 1,086 | 1,075 |
| R-squared | 0.380 | 0.382 | 0.384 | 0.388 |
| R-squared adj. | 0.377 | 0.375 | 0.377 | 0.377 |
| Mean(dependent variable) | 0.078 | 0.078 | 0.078 | 0.078 |

*Note:* p-values in parentheses: * $p < 0.10$, ** $p < 0.05$, *** $p < 0.01$. Estimated using ordinary least squares. Standard errors are clustered at the district level.